\documentstyle[12pt]{article}
\input{epsf}
\begin{document}
%
%%%%%%%%%%%%%%%%%%%%%%%%%%%%%%%%% definitions %%%%%%%%%%%%%%%%%%%%%%%%%%%%%%%%%
\newcommand{\nc}{\newcommand}
\def\veffT{\Delta V_{{\rm eff},T}}
\def\msh{m^2_{h^0}}
\def\veffo{V_{{\rm eff},0}}
\nc{\beq}{\begin{equation}}
\nc{\eeq}{\end{equation}}
\nc{\beqa}{\begin{eqnarray}}
\nc{\eeqa}{\end{eqnarray}}
\nc{\lra}{\leftrightarrow}
\nc{\sss}{\scriptscriptstyle}
{\nc{\lsim}{\mbox{\raisebox{-.6ex}{~$\stackrel{<}{\sim}$~}}}
{\nc{\gsim}{\mbox{\raisebox{-.6ex}{~$\stackrel{>}{\sim}$~}}}
\def\dsl{\partial\!\!\!/}
\def\lameff{\lambda_{\rm eff}}
\def\Re{{\rm Re\,}}
\def\Im{{\rm Im\,}}
\def\diag{{\rm diag}}
\def\sign#1{{\rm sign}(#1)}
\def\VEV#1{{\langle #1 \rangle}}
%234567890123456789012345678901234567890123456789012345678901234567890123456789
%%%%%%%%%%%%%%%%%%%%%%%%%%%%%%%%%%%%%%%%%%%%%%%%%%%%%%%%%%%%%%%%%%%%%%%%%%%%%%%
\begin{titlepage}
\pagestyle{empty}
\baselineskip=21pt
\rightline{McGill 97-26}
\rightline{HIP--1997-44/TH}
\rightline{hep-ph/9708393}
\rightline{August, 1997}
\vskip .4in
\begin{center}
{\Large{\bf 
Supersymmetric Electroweak Baryogenesis \\
in the WKB Approximation}}
\end{center}
\vskip .1in
\begin{center}
James M.~Cline\\
{\it McGill University, Montr\'eal, Qu\'ebec H3A 2T8, Canada}\\
Michael Joyce \\ 
{\it School of Mathematics, Trinity College, Dublin 2, Ireland}\\
Kimmo Kainulainen\\
{\it  High Energy Physics Division,\\
P.O.\ Box 9, FIN-00014, University of Helsinki 
}
\vskip .2in
\end{center}

\vskip 0.7in \centerline{ {\bf Abstract} } 
\baselineskip=18pt 
\vskip
0.5truecm 

We calculate the baryon asymmetry generated at the electroweak phase
transition in the minimal supersymmetric standard model, treating the
particles in a WKB approximation in the bubble wall background.  A set
of diffusion equations for the particle species relevant to baryon
generation, including source terms arising from the CP violation
associated with the complex phase $\delta$ of the $\mu$ parameter, are
derived from Boltzmann equations, and solved.  The conclusion is that
$\delta$ must be $\gsim 0.1$ to generate a baryon asymmetry consistent
%%%%%%%%%%%%%%%%%%%%%%%%%%%%%%%%%% REWORDED below %%%%%%%%%%%%%%%%%%%%%%%
with nucleosynthesis.  We compare our results to several other recent
computations of the effect, arguing that some are overestimates.
%%%%%%%%%%%%%%%%%%%%%%%%%%%%%%%%%% REWORDED above %%%%%%%%%%%%%%%%%%%%%%%
\end{titlepage} 
\newpage

\baselineskip=20pt

Although theories to explain the baryon asymmetry of the Universe (BAU)
abound, only electroweak baryogenesis has the hope of short-term
testability. To calculate the asymmetry in detail one needs a specific
model of electroweak physics beyond the standard model, and
supersymmetry is the most popular idea for such new physics.  Several
groups have recently estimated the BAU coming from the minimal
supersymmetric standard model (MSSM) at the electroweak phase
transition (EWPT), assuming it is first order and so proceeds by bubble
%%%%%%%%%%%%%%%%%%%%%%%%%% added reference DRW %%%%%%%%%%%%%%%%%%%%%%%
nucleation \cite{HN}-\cite{DRW}. Despite the fact that the physics of the
mechanism is essentially agreed upon---particles in the plasma interact
with the bubble wall in a CP-violating manner, leading to a chiral
asymmetry of quarks in front of the wall, which in turn biases
sphalerons to produce the BAU---there is considerable disagreement on
both methods and results.

While it is straightforward to formulate diffusion equations for
particle transport away from the wall, it is less obvious how to derive
the CP-violating source terms (or equivalently the boundary conditions
at the wall) for these equations, and it is here that most of the
difficulty lies. Several methods have been proposed in which the source
terms are first derived and then inserted into a set of diffusion
equations, but without any first principles justification for the
prescription used.  In the present work we generalize a method which
was introduced for the treatment of the non-supersymmetric two doublet
model in \cite{jpt}, in which the diffusion equations and the source
terms are {\it derived together} starting from a general set of
classical Boltzmann equations.  This provides a systematic and
controlled approximation valid for bubble walls significantly thicker
than the inverse temperature, which is the case in the MSSM.  Here we
will describe the results as concisely as possible; finer details will
be presented in a longer publication \cite{next}.

The method is based on a WKB quasi-particle approximation to the
particle dynamics and interactions in the background of the bubble
wall, which is an expansion in derivatives of the Higgs field VEV's
that vary continuously in the wall. Such an expansion is appropriate
because it well-describes the majority of particles in the plasma,
whose wavelength ($\sim 1/T$) is much shorter than the thickness of the
wall ($\sim 20/T$). When there is CP violation in the theory, such as
is provided by the complex $\mu$ and $A_t$ parameters of the MSSM, this
approximation shows that particles and anti-particles experience a
different force when interacting with the bubble wall, both dynamically
as free WKB particles and in their interactions with other particles in
the plasma.  These two effects---the classical dynamical force on the
particles, and the CP biasing of their collisions---appear in the
Boltzmann equations for the WKB particles in a quite straightforward
way, through the force term and collision term, respectively.  These
equations determine the distribution of left-handed quark number
($q_L$) in front of the bubble wall, which is the source term for
baryon number generation.  An appropriate fluid-type truncation leads
to a set of diffusion equations of a familiar form, together with the
source terms, which predict the $q_L$ and hence the baryon asymmetry.

In what follows we will concentrate on charginos as the source of the
chiral quark asymmetry, because we find they make the dominant
contribution.  The corresponding analysis for the subdominant squark
and quark contributions will be presented in ref.~\cite{next}.  To find
the effect of CP violation on charginos at the bubble wall, we first
solve the Dirac equation in the WKB approximation.

{\bf WKB approximation for charginos.}
The mass term for charginos can be written
as $\overline\Psi_R M_\chi \Psi_L +$ h.c., where in the basis of Winos
and Higgsinos, $\Psi_R = (\tilde W^+_R, \tilde h^+_{1,R})^T$, $\Psi_L =
(\tilde W^+_L, \tilde h^+_{2,L})^T$ the mass matrix is
\beq
        M_\chi = \left(\begin{array}{cc} m_2 & g v_2/\sqrt{2} \\
                        g v_1/\sqrt{2} & \mu \end{array}\right),
\label{chmm}
\eeq
with the spatially varying Higgs field VEV's $v_i$. The corresponding 
Dirac equation is $
(i\gamma_3\gamma_0 \partial_t  + i \partial_z - \gamma_3 ( M_\chi P_L
        + M_\chi^\dagger P_R))\Psi = 0$.  
In the WKB approximation the local eigenstates of mass, energy and helicity
are related to the flavor states through a transformation 
\beq
	\left(\begin{array}{c} \Psi_L \\ \Psi_R \\ \end{array}\right) = 
	\left(\begin{array}{cc} V & 0\\ 0 & U \\ \end{array}\right) 
	\left(\begin{array}{cc} \cosh(X)
        & e^{i\theta}\sinh(X)\\ e^{-i\theta}\sinh(X) & \cosh(X)
        \end{array}\right)
	\left(\begin{array}{c} \Psi_{\downarrow} \\ 
	\Psi_\uparrow \\ \end{array}\right)
	e^{-iEt + i\int^z p(z) dz}.
\label{transf}
\eeq
Here $U$ and $V$ are the unitary transformations that locally
diagonalize the mass matrices, $U^\dagger M_\chi V= {\cal M}_\chi$ and
$V^\dagger M^\dagger_\chi U = {\cal M}^*_\chi$, and $X$ is defined by
$\tanh(2X) = |{\cal M}_\chi|/E$, which determines the mixing between
the chirality and the helicity states.  $\theta$ is a diagonal matrix,
consisting of the phases of the eigenvalues ${\cal M}_\chi$, whose
spatial variation is responsible for all the CP violating effects we
will be concerned with.  Finally, $p(z)$ is also a diagonal matrix that
gives the local momentum of the two mass eigenstates.   The explicit
expressions for the mass eigenvalues, unitary matrices and CP-violating
phases can be given in terms of the quantities
\beqa
\label{deltaeq}
 	\tilde m^2 &=& (m_2^2 + |\mu|^2 + u_1^2 + u_2^2)/2; \qquad
	u_i = g v_i/\sqrt{2}; \nonumber\\
        \Delta &=& (m_2^2 - |\mu|^2 - u_1^2 + u_2^2)/2; \qquad
        a = m_2 u_1 + \mu^* u_2 \nonumber \\
        \bar\Delta &=& (m_2^2 - |\mu|^2 + u_1^2 - u_2^2)/2; \qquad
        \bar a  = m_2 u_2 + \mu u_1 \nonumber \\
        \Lambda &=& (\Delta^2 + |a|^2)^{1/2} = (\bar\Delta^2 + |\bar
        a|^2)^{1/2};
\eeqa
then $|{\cal M}_{\chi,\pm}|^2 = \tilde m^2 \pm \Lambda$, and
\beq
        U = {1\over\sqrt{2\Lambda(\Delta+\Lambda)}}
        \left(\begin{array}{cc} \Delta + \Lambda & -a \\
        a^* & \Delta + \Lambda \end{array}\right),\qquad 
        V = {1\over\sqrt{ 2\Lambda(\bar\Delta+\Lambda)}}
        \left(\begin{array}{cc} \bar\Delta + \Lambda & -\bar a 
        \\ \bar a^* & \bar\Delta + \Lambda \end{array}\right),
\label{UandV}
\eeq
and 
\beq
	|{\cal M}_{\chi,\pm}|^2 {\partial\theta_\pm} = \pm g v_c \Im(\mu)
	m_2 \cos\beta\sin\beta u\partial_z u / \Lambda,
\eeq
where $\partial_z$ denotes the spatial derivative normal to the bubble
wall, $v_c = \sqrt{v_1^2+v_2^2}$ is the magnitude of the VEV at the
critical temperature, $u=\sqrt{u_1^2+u_2^2}$ and $\tan\beta =
v_2/v_1$.  In what follows the phase $\theta$ will never be needed
except in the combination shown above.  Finally, to lowest nontrivial
order in derivatives and CP-violation, the local momenta are $p(z)  =
\sign{p_z}\left(\sqrt{E^2-|{\cal M}_{\chi}|^2}
\mp\sinh^2(X)\partial_z\theta\right)$, for particles with helicity
$\pm$, and in the Lorentz frame where the particle has no momentum
parallel to the bubble wall.  For the antiparticles, the same equation
holds but with an overall change of sign for the CP-violating part.

The important thing to notice in these expressions is that the
CP-violating angle $\theta$ appears in two places: in the particle
momenta (as a derivative) and in the transformation matrices to the WKB
basis. The former leads to the dynamical force effect and the latter to
source terms of the ``spontaneous'' type, which have previously been
discussed in the literature in various contexts \cite{sbg}. In what
follows it will be convenient to have the dispersion relation for
energy in terms of momentum, for the two mass eigenstates $i$:
\beqa
\label{Efermion}
        E_{\pm} &=& \sqrt{\vec p^{\,2}+|m_i|^2} \pm \sign{p_z}
         {\partial_z \theta_i\over 2
        \sqrt{\vec p^{\,2}+|m_i|^2}}\left( \sqrt{p_z^{2}+|m_i|^2}-|p_z|
        \right) \equiv  E_0 \pm \Delta E;\nonumber\\
       \pm &=& +\hbox{\ for\ }R,\bar L,\ -\hbox{\ for\ }L,\bar R.
\label{dispeq}
\eeqa
To derive this we used the identity
$\sinh^2(X)={\left(\sqrt{p_z^{2}+|m|^2}-|p_z| \right)/(2|p_z|)}$ and
transformed to a general Lorentz frame with nonzero momentum parallel
to the wall.

{\bf Diffusion equations with sources.}  
Our starting point is the Boltzmann equation 
 $(\partial_t +
\dot{\vec{x}}\cdot\partial_{\vec{x}}  +
\dot{\vec{p}}\cdot\partial_{\vec{p}}) f_i = C[f_i]$ for the local
distributions $f_i(\vec{p}, \vec{x})$ of the WKB quasi-particles 
with the on-shell dispersion relation just derived. The on-shell
approximation is valid since the particle widths $\Gamma$ are
small compared to their energies. To truncate these
equations we use the ansatz  
\beq
        f_i (p,x) = {1\over e^{\beta(E_i - v_i p_z - \mu_i)} \pm 1},
\label{dist}
\eeq
for the distribution functions in the rest frame of the plasma, where
$E_i$ contains the perturbation $\Delta E_i \sim \partial\theta$, {\it
i.e.,} we allow a chemical potential and velocity perturbation in each
fluid. The validity of the ansatz requires that $v_w < L/3D$
\cite{jpt}, in order that whatever other perturbations are produced are
indeed small compared to those of the chemical potential, the latter
being damped only by the slower inelastic processes.  The velocity
perturbation is required to model the anisotropic response to the force
term.  Expanding to linear order in the perturbation $\Delta E_i$ and
taking $\mu_i$ and $v_i$ to be of this order gives $f_{\pm}'
\partial_z  \Bigl( (v_w-p_z/E_0) ( p_z v_i + \mu_i) - v_w  \Delta E_i
\Bigr) = - C\left[f_i \right]$, where $f_{\pm}'=df_{\pm}/dE$ denotes
the derivative of the unperturbed Fermi-Dirac or Bose-Einstein
distribution,  $f_{\pm}'=(e^{\beta E_0} \pm 1)^{-1}$.  Then averaging
over momentum, weighting the equation respectively by $1$ and $p_z$, we
obtain the two coupled equations %for $\mu_i$ and $v_i$:
\beqa
\label{diffeq1}
        -\VEV{p_z^2/E_0}v_i' + v_w \mu_i' &=& \VEV{C_i};\\
\label{diffeq2}
        v_w\VEV{p_z^2} v_i' - \VEV{p^2_z/E_0}\mu_i' - v_w\VEV{p_z \Delta
        E_i'} &=&  \VEV{C_i p_z}
\eeqa
where $\VEV{\cdot}\equiv \int d^3 p f_\pm'(\cdot)/\int d^3 p f_+'
\equiv \kappa_i \int d^3 p f_\pm'(\cdot)/\int d^3 p f_\pm'$ and the
primes in eqs.~(\ref{diffeq1}-\ref{diffeq2}) denote $\partial_z$.  Here
we have divided by the distribution function for a massless fermion
even if the numerator is for a boson, because this ensures that the
collision integral on the r.h.s.~of the Boltzmann equation gives the
same value for the rate of a given process, regardless of whether one
is writing the transport equation for a boson or for a fermion
involved in that process.  The result of this choice is to introduce
the factor of $\kappa_i$ on the l.h.s.~of the Boltzmann equation, which
is 1 for fermions and 2 for bosons, in the massless approximation.

To evaluate the collision terms in eq.~(\ref{diffeq1}-\ref{diffeq2})
one must substitute the distributions (\ref{dist}) into $\int d\Pi_j
\delta^{(4)}(p_i+p_2\dots-p_{n-1}-p_n) |{\cal M}|^2(f_i f_2\cdots (1\pm
f_{n-1})(1\pm f_n) - f_n f_{n-1}\cdots (1\pm f_2)(1\pm f_i))$, which
vanishes for the equilibrium distibutions.  Thus the collision terms
are linear in the perturbations: $\VEV{C_i}= \sum_j\Gamma^d_{i\to j}
\mu_j$ and $\VEV{p_z C_i} = \VEV{p_z^2}\sum_j\Gamma^e_{ij\to ij}v_i$,
%(v_i-v_j)$, 
where $\Gamma^{d,e}$ are respectively the inelastic
(decays) and elastic interaction rates for processes coupling states
$i$ and $j$.  There is further another important contribution to
$\VEV{C_i}$ coming from the fact that the interactions of the flavor
eigenstates, when reexpressed in terms of the local mass eigenstates,
have factors of $e^{i\theta(z)}$ coming from eq.~(\ref{transf}).  In
fact that equation is derived in the rest frame of the bubble, so that
in the wall frame $e^{i\theta(z)}\to e^{i\theta(z-v_wt)}$, which to
leading order in derivatives can be written as $e^{i(\theta(z)-
v_w\partial\theta)}$.  The effect is to spoil energy conservation of
the interaction so that $\delta(E_i+E_2\dots-E_{n-1}-E_{n}) \to
\delta(E_i-v_w\partial\theta+E_2\dots-E_{n-1}-E_{n})$.  This gives an
additional term in $\VEV{C_i}\sim v_w\partial\theta$, which is of the
type referred to as a spontaneous baryogenesis source term.   
In the case of Higgsinos, the interaction $y
\bar q_L \tilde t_R \tilde h_2$ can be rewritten as $y \bar q_L \tilde
t_R (V_{2+} \sinh(X_+) e^{i\theta_+} \Psi_{\uparrow+} + V_{2-}
\sinh(X_-) e^{i\theta_-}\Psi_{\uparrow-})$ leading to $\VEV{C_{\tilde
h_2}} = v_w\Gamma_y|V_{2\pm}|^2
\VEV{\sinh^2(X_\pm)}\partial\theta_{\pm}$, where $\pm$ is whichever
sign is needed to make the associated mass eigenstate go smoothly into
the pure Higgsino state in front of the wall (see below).

Finally, the two coupled equations can be reduced to a single one
by differentiating (\ref{diffeq2}) and eliminating $v_i$ in favor
of $\mu_i$.  Defining the diffusion constant 
$D_i =\VEV{p_z^2/E}^2/(\VEV{p_z^2} \Gamma^e_i)$ and ignoring the ratio
of inelastic to elastic scatterings, we obtain the diffusion equation
\beqa
&&     -\kappa_i( D_i\mu_i''+ v_w \mu_i')+\Gamma^d_{i} \sum_j \mu_j = S_i;
        \nonumber\\
&&      S_i = {D_i v_w\over\VEV{p_z^2/E}}\VEV{p_z\Delta E_i'}'
        -v_w\sum_j\Gamma^d_{j\to i}|V_{ji}|^2\VEV{\sinh^2(X_i)}\theta'_i,
\label{diffeq3}
\eeqa
to leading order in the wall velocity.

{\bf Solution of diffusion equations.}  
{}From this complicated network of equations coupling all species in
the plasma we need to determine the distribution of left-handed 
fermions, which bias the unsuppressed baryon-number-violating processes
in front of the wall. It can be simplified by using conservation laws
and neglecting processes which are too slow to play any role
on the relevant timescales.  Parametrically this means comparing
the time spent by a particle diffusing in front of the wall $\sim D/v_w^2$
with the decay time $\Gamma^{-1}$. 
The processes we do take account of are then (i) the supergauge interactions,
(ii) those described by the following terms in the interaction Lagrangian
\beqa
   V_y = y h_2\bar u_R q_L + y \bar u_R \tilde h_{2L}\tilde q_L 
   + y \tilde u^*_R \tilde h_{2L} q_L
   - y \mu h_1 \tilde q^*_L \tilde u_R + yA_t\tilde q_L h_2\tilde u^*_R
   +\hbox{h.c.}
\label{Vint}
\eeqa
and (iii) strong sphaleron interactions. 
Taking the supergauge interactions to be in equilibrium,
the chemical potentials of all gauginos are zero and 
the chemical potential of any particle is equal to that of
its superpartners. Defining a reduced set of chemical potentials 
for each generation and chirality of baryons and each Higgs doublet, 
$H_{1,2}$, $Q_{1,2,3}$, $U,D,S,C,B,T$, with 
$H_1=\frac14(\mu_{h^0_1}+\mu_{h^-_1}
+\mu_{{\tilde h}^0_1}+\mu_{{\tilde h}^-_1})$, 
$Q_1=\frac14(\mu_{u_L}+\mu_{d_L}+\mu_{{\tilde u}_L}+\mu_{{\tilde d}_L}),$
$U=\frac12(\mu_{u_R}+ \mu_{\tilde{u}_R}),$ etc.,
the interaction terms are
\beqa
&&(\Gamma_{y}+\Gamma_{yA})(-H_2+Q_3-T),\quad \Gamma_{y\mu}(H_1-Q_3+T),	
\quad \Gamma_{hf}(H_1+H_2),\nonumber\\
&&
\quad \Gamma_{ss}(2Q_3 + 2Q_2 + 2Q_1 - U-D-C-S-B-T).
\label{gammas}
\eeqa
Further conservation laws for various unsourced linear combinations
give us that $Q_2=Q_1$ and $U=D = S = C = B$.  Now putting   
all the Yukawa interactions in equilibrium gives the constraints 
$-H_2+Q_3-T=0$ and $H_1-Q_3+T = 0$, and using the conservation
of total baryon number (the weak sphalerons are slow and enter
only at the end of the calculation) allows the full network
to be reduced to just two coupled equations for $Q_3$  
and $H\equiv H_1+H_2$:
\beqa
        -3{\cal D}_h H - 3{\cal D}_q Q_3 +\Gamma_{hf}H +
        \Gamma_{ss}(28 Q_3-4H) &=&  S_{H}\equiv \frac12
        (S_{H_1}+S_{H_2})\nonumber\\
        +\frac32{\cal D}_q H - 9{\cal D}_q Q_3 + \Gamma_{ss}(28 Q_3-4H),
        &=&0.
\label{final-two-eqns}
\eeqa
where ${\cal D}_i\equiv D_i{d^2\over d z^2} + v_w{d\over dz}$.
We must keep the strong sphaleron and helicity-flip rates explicitly
because in the limit that either becomes infinitely fast, they drive
the left-handed quark number, essential for baryogenesis, to zero. The
Yukawa interactions, by contrast, simply redistribute the asymmetries
between the species, transferring the chemical potential from
Higgsinos, where the source is, into the left-handed quark sector.  It
is well known that the strong sphalerons tend to drive the asymmetry to
zero in the massless approximation we are working in \cite{GS} and,
since $B_L = Q_1 + Q_2 + Q_3 = 14Q_3 - 2H$, this is manifest in
eq.~(\ref{final-two-eqns}). For the helicity flips, which are due to
the Dirac mass term $\mu \tilde h_1 \tilde h_2$ that connects the two
species of Higgsinos, the potential suppression arises from the fact
that $S_{H_1} = S_{H_2}$, since  $\tilde h_1$ and $\tilde h_2^c$ are the
right- and left-handed components of a Dirac fermion, and thus come
with equal and opposite CP phases, as can be seen in
eq.~(\ref{dispeq}).  If helicity-flipping interactions were in
equilibrium this would force $H_1=-H_2$ and the source for $H_1-H_2$ would 
be $S_{H_1}- S_{H_2}=0,$ leading to the trivial solution
for all the particle densities.\footnote{This statement is true for the
force-term contribution to $S_H$, but the unequal interactions of $H_1$
and $H_2$ prevent the spontaneous baryogenesis part of $S_{H_1}-
S_{H_2}$ from vanishing.  In practice however it is small.}

The equations (\ref{final-two-eqns}) can be solved by finding the
appropriate Green's function, of the form ${\bf G}(z-z') = \int {\bf
G}(p) e^{ip\cdot(z-z')} dp/2\pi$, which is a matrix such that
$(H(z),Q_3(z))^T =  \int_{-\infty}^\infty {\bf G}(z-z')(S_H(z'),0)^T
dz'$.  However, we do not need the full matrix $\bf G$, because we are
only interested in the linear combination of $Q_3$ and $H$ that gives
the chemical potential for left-handed baryon number, $B_L  = 14Q_3 -
2H$.  The latter can be determined from the single-component equation
\beq
	B_L(z) = \int_{-\infty}^\infty {G_B}(z-z')
	S_{H}(z') dz',
\label{BLeq}
\eeq
where $G_B = -2{\bf G}_{11} + 14{\bf G}_{21}$.  Carrying out the
thermal averages in the relativistic limit (for simplicity of 
presentation--we do not make this simplification to obtain the numerical
results),
and using the Maxwell-Boltzmann
approximation for the distribution functions, the source can be written
as
\beq
	S_H = v_w %e^{-m/T}P(m/T)
	\left({D_h\over 4\VEV{p_z^2/E}}
	\left((m^2\theta')'
	\VEV{1/E}\right)' - {\Gamma_y\over 8}\VEV{1/E^2}
	m^2\theta'\right),
\eeq
where to a good approximation %$P(x) = 1+1.1x + 0.25x^2$,
$\VEV{p_z^2/E} = T$, $\VEV{1/E} = 1/(2T+m)$, and $\VEV{1/E^2} =
(T+m)^{-1}(2T+m)^{-1}$.  The appropriate mass $m$ to use here is
$|{\cal M}_{\chi\pm}|$ (see above eq.~(\ref{UandV})), with $\pm$ equal to
$\sign{\mu-m_2}$: this is the appropriate sign for the mass eigenvalue
that goes continuously into that of the chargino ($m=\mu$) in the 
symmetric phase outside the bubble.

The Green's function is determined by the poles of $G_B(p)$, which are
the roots of a quartic polynomial.  For $v_w <  \sqrt{\Gamma D}$ we can
solve perturbatively in $v_w$, and the result to $O(v_w)$ is
\beqa
  G_B(z) = {\cal C} \sum_\pm \pm k_\pm e^{-k_\pm|z|}\left(1+v_w(\sign{z}
   \delta g_\pm  - \delta k_\pm z)\right) +O(v_w^2) 
\label{GBeq}
\eeqa
where
\beqa
 k^2_\pm &=& {28 \Gamma_{ss} D_h+3(2\Gamma_{ss}+\Gamma_{hf})D_q \pm
	\sqrt{(28\Gamma_{ss} D_h +3(2\Gamma_{ss} -\Gamma_{hf})D_q)^2 
+16\Gamma_{hf}\Gamma_{ss} D_q^2)}\over 18D_q(D_h+D_q) } \nonumber \\
\delta  k_\pm &=& { 3(3D_h+4D_q)k^2_\pm -(3\Gamma_{hf}+34\Gamma_{ss})
\over 6D_q(6D_h+D_q)k^2_\pm -56 \Gamma_{ss} D_h - 
6(\Gamma_{hf}+2\Gamma_{ss}) D_q }
\eeqa
and  ${\cal C}^{-1} = 3(6 D_h+D_q)(k_+^2-k_-^2)$, $\delta g_{\pm} =
(2\delta k_{\pm}-1/D_q)/k_{\pm}- 2k_{\pm}(\delta k_+ - \delta
k_-)/(k_+^2-k_-^2) $.  It is important to go to first order in $v_w$
here because of a cancellation that takes place at $O(v_w^0)$: the
Green's function at this order is symmetric about the center of the
wall, but the source is approximately antisymmetric.

{\bf The baryon asymmetry.}  Having the solution for the density of
left-handed baryon number, it is simple to find the rate of baryon
violation due to weak sphalerons in front of the wall: $ \dot n_B =
-9\Gamma_{ws} n_{q_L} = -9\Gamma_{ws} T^2 B_L$.  The time integral of
this rate from $t=-\infty$ until the time the wall passes a
given position, where the sphaleron interactions are assumed
to switch off, can be converted to an integral of $B_L(z)$ in front of 
the wall, yielding 
the baryon density $n_B = (9\Gamma_{ws} T^2 /v_w)\int_0^\infty 
B_L(z) dz$.  The baryon-to-photon ratio is therefore
\beq
	\eta \cong {7 n_B\over s} = {2835\alpha_W^4\over 2\pi^2 v_w g_*}
	\int_0^\infty B_L(z)\, dz
\eeq
where $g_*$ is the number of degrees of freedom at the phase transition
temperature.  To compute $\int B_L(z) dz$, we use eq.~(\ref{BLeq}) and
(\ref{GBeq}), where the $z$ integral can be done analytically, leaving
a single numerical integral over $z'$.

To obtain the final result we must still specify the shape and speed of
the bubble wall and the values of diffusion constants and interaction
rates.  For the wall profile, we took the kink solution $v(z) = (v_c/2)
\left(1-\tanh(z/2w)\right)$ with width $w = 1/(\sqrt{\lambda} v_c)$,
corresponding to the Higgs potential at the critical temperature $V(v)
= \lambda v^2 (v-v_c)^2$.  We explored a range of values of $\lambda$
and $v_c$ consistent with the analysis of the phase transition
described in refs.~\cite{ck}, giving wall widths between $w=0.07$ and
$w=0.08$ GeV$^{-1}$, which should be compared to the typical critical
temperature of $T_c=90$ GeV.  The other parameters varied in the ranges
$0.003 < \lambda < 0.014$ and $120$ GeV $< v_c < 210$ GeV. For the
diffusion constants and decay rates, we take $\Gamma_{ss}= 6000
\Gamma_{ws}$ \cite{moore}, $\Gamma_{ws}=\alpha_w^4 T$ (it has recently
been argued that parametrically $\Gamma_{ws}=C \alpha_w^5 T$
\cite{ASY}, but lattice measurements of the rate are consistent with a
value of $C\sim 1/\alpha_w$ \cite{mt} so this does not effect the
numerical value of our estimate) and we have made the rough
estimates  $\Gamma_{hf} = 3y^2 T/16\pi$, $D_h = 20/T$ and $D_q = 3/T$.
Our results are not extremely sensitive to these values.  For the wall
velocity we took $v_w = 0.01-0.2$.

In general, we find that the baryon-to-photon ratio $\eta$ is
sufficiently large only if the CP-violating phase $\delta$ of the $\mu$
parameter is nearly maximal, $\delta = \pi/2$.  In figure 1 we show the
dependence of $\eta_{10} = \eta\times 10^{10}$ on the chargino mass
parameters $m_2$ and $\mu$ for $\tan\beta=2$ (favored by studies of the
phase transition) $\lambda = 0.0065$, $v_c = 160$ GeV, and $w=7/T_c$,
and several values of the wall velocity.  One finds certain parameter
values, like  $m_2\sim\mu$ when $v_w\gsim 0.1$, where a smaller angle
of $\delta\sim 0.1$ could be tolerated.  It is usually assumed that
such a large phase is excluded by the experimental limits on the
neutron electric dipole moment, but this can be evaded if the top and
bottom squarks are much lighter than their first generation
counterparts, since these determine the size of the loop diagrams that
generate the up and down quark EDM's that feed into that of the
neutron.  This is in some sense natural because a light top squark is
also needed for a strong enough phase transition to satisfy the
sphaleron bound.

%%%%%%%%%%%%%%%%%%%%%%%%%%%%%%%%%%%%%%
%%%  figure 1
%%%%%%%%%%%%%%%%%%%%%%%%%%%%%%%%%%%%%%
\begin{figure}
\epsfysize = 1.5in
\centerline{\epsfbox{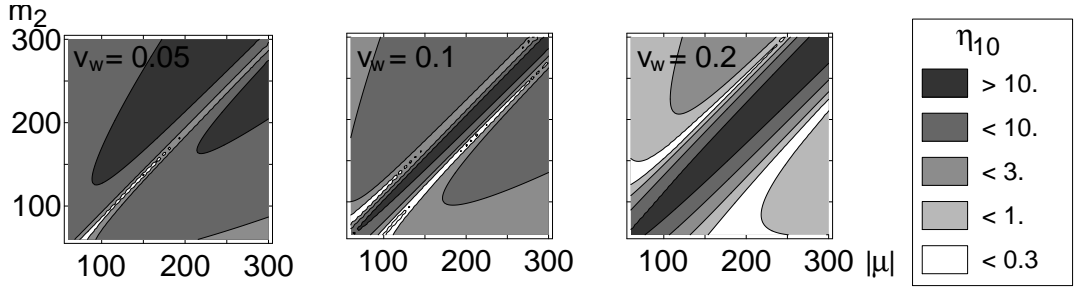}}
\baselineskip=16pt
\caption{Contours of constant $\eta_{10}=10^{10}\eta$ in the plane of 
$m_2$ (the Wino mass) and $\mu$, in GeV, $v_{\rm wall}= 0.05$, 
$0.1$ and
$0.2$, $\tan\beta=2$, $\sin\delta=1$ and wall width $2w = 14/T$.
}\end{figure}
\vskip 0.25in \baselineskip=20pt

{\bf Discussion.}  We conclude by pointing out some of the important
differences between previous related work and ours, and why we consider
the present analysis to be more satisfactory.  The comparison can be
divided into two parts: the derivation of the sourced diffusion
equations, and their subsequent solution and estimation of the baryon
asymmetry. With respect to the first point we have noted that in the
present treatment the source terms and diffusion equations are
consistently derived together.  In several recent treatments
\cite{HN,CQRVW,Worah}, quantum mechanical contributions to various
CP-odd currents induced by the wall are derived, and then converted
into sources for diffusion equations using an ad hoc prescription
introduced in \cite{HN}.  If such a prescription could be rigorously
derived, the derivation should start with a set of transport equations
more general than those used here, incorporating quantum mechanical
corrections.  However any such consistent treatment should reduce to
our equations in the classical limit, and this is not the case, as can
be seen from the fact that the source terms of \cite{HN,CQRVW,Worah}
are parametrically different from ours.  Furthermore, insofar as the
thick wall and on-shell quasi-particle approximations we have used are
valid, such quantum mechanically derived source terms should be higher
order corrections to those derived here.

One example of the different parametric dependence of our source term
concerns the variation of $\tan\beta$ in the wall.  The source terms
discussed in refs.~\cite{HN,CQRVW,AOS2,DRW} are proportional to $v_1 v_2' -
v_1' v_2$, which is zero if $v_1/v_2$ is constant in the bubble wall.
As pointed out in ref.~\cite{CQRVW} and confirmed by the methods of
ref.~\cite{ck}, $v_1/v_2$ varies at most by only a few percent over the
wall in the MSSM.  In ref.~\cite{AOS2}, on the other hand, it is simply
assumed that $v_1/v_2$ changes by 100\% in the wall, which gives an
unjustifiably large estimate of the baryon asymmetry.  We have
concentrated on the chargino source because, in our derivation, it is
apparently the sole exception to this rule in the MSSM.  It escapes the
suppression because the chargino is a Dirac particle, whose mass
eigenvalues have a spatially varying phase when $v_1/v_2$ is constant,
even though the eigenvalues of $M^\dagger M$ have no such
phases.\footnote {Actually the suppression would still occur if Winos
and Higgsinos had identical interactions, since they come with opposite
phases.  It is the different interaction rates of the two species that
undoes the suppression coming from constant $v_1/v_2$.}

%%%%%%%%%%%%%%%%%%%%%%%%%%%% REWORDED below%%%%%%%%%%%%%%%%%%%%%%%%%%%%%% 
There are several other differences between our treatment and previous
ones.  First, our computation has assumed that all the squarks are
light and hence the strong sphaleron suppression must be taken into
account, whereas \cite{CQRVW,Worah} took all squarks except for $\tilde
t_{L,R}$ and $\tilde b_L$ to be decoupled.  In this case the strong
sphaleron suppression is evaded \cite{HN}.  
Secondly we find that for $v_w > 0.01$ (so
that electroweak sphalerons are still out of equilibrium on the wall
passage time scale), the baryon asymmetry is $O(1)+ O(10 v_w)$ rather
than $O(1/v_w)$ as found by refs.~\cite{HN,Worah}.\footnote{We have
been informed by the authors of ref.~\cite{CQRVW} that their results
are consistent with no velocity dependence, despite the apparent
$O(1/v_w)$ behavior which seems to be indicated by their equations.}
%%%%%%%%%%%%%%%%%%%%%%%%%%%%%%%%%% REWORDED above %%%%%%%%%%%%%%%%%%%%%%%
This can be traced to the fact that the Higgsino decay rate is taken to
be zero in the symmetric phase, but non-zero inside the bubbles.  In
fact we believe it should be the other way around since Higgsinos have
much more phase space to decay into massless quarks and light stops in
the symmetric phase than in the broken phase where the top quark is
heavy.  But regardless of phase space, helicity-flipping scattering
processes are still fast enough in the symmetric phase to give a
significant rate of Higgsino damping.

{\bf Acknowledgement.}  MJ is supported by an Irish Government (Dept.~of
Education) post-doctoral fellowship.


\begin{thebibliography}{99}
\bibliographystyle{unsrt}
\baselineskip 13 pt

\bibitem{HN}P.~Huet and A.E.~Nelson, hep-ph/9506477, Phys.~Rev.~{\bf
D53},{4578} (1996).
\bibitem{AOS1}M.~Aoki, N.~Oshimo and A.~Sugamoto, hep-ph/9612225 (1996).
\bibitem{CQRVW}M.~Carena, M.~Quiros, A.~Riotto, I.~Vilja and C.E.M.~Wagner,
preprint CERN-TH-96-242, hep-ph/9702409 (1997).
\bibitem{Worah}M.P.~Worah, hep-ph/9702423, 
	to appear in Phys.~Rev.~{\bf D56} (1997). 
\bibitem{AOS2}M.~Aoki, N.~Oshimo and A.~Sugamoto, hep-ph/9706287 (1997).
%%%%%%%%%%%%%%%%%%%%%%%%%% added reference DRW %%%%%%%%%%%%%%%%%%%%%%%
\bibitem{DRW} H.~Davoudiasl, K.~Rajagopal and E.~Westphal, Caltech
	preprint CALT-68-2127, hep-ph/9707540 (1997).
\bibitem{jpt}
 M.~Joyce, T.~Prokopec and N.~Turok, Phys.~Lett.\
{\bf B338}, 269 (1994); Phys.\ Rev.\ Lett.\ {\bf 75},1695 (1995), 
ERRATUM-{\it ibid.}~{\bf 75}, 3375 (1995);  Phys.~Rev.~{\bf D53}, 2930 (1996);
Phys.~Rev.~{\bf D53}, 2958 (1996).
\bibitem{next}
J.~Cline, M.~Joyce and K.~Kainulainen, in preparation (1997).
\bibitem{sbg} A.G.~Cohen and D.B.~Kaplan, Phys.~Lett.~{\bf B199}, 257 (1987);
Nucl.~Phys.~{\bf B308}, 913 (1988);\\
 A.G.~Cohen, D.B.~Kaplan and A.E.~Nelson,  Phys.~Lett.~{\bf B336},41 (1994).
\bibitem{GS}  G.F.~Giudice and M.~Shaposhnikov, Phys.~Lett.~{\bf B326},
118 (1994).
\bibitem{ck} J.M.~Cline and K.~Kainulainen, Nucl.~Phys.~{\bf B482},
73-91, hep-ph/9605235 (1996); hep-ph/9705201 (1997).
\bibitem{moore} G.D.~Moore, preprint PUPT-1698, hep-ph/9705248 (1997).
\bibitem{ASY} P.~Arnold, D.~Son and L.G.~Yaffe,  hep-ph/9609481,
Phys.~Rev.~{\bf D55}, 6264 (1997). 
\bibitem{mt} G.D.~Moore and N.~Turok, preprint PUPT-1681, hep-ph/9703266
(1997).

\end{thebibliography}
\end{document}